\documentclass[preprint, showpacs,preprintnumbers,amsmath,amssymb,nofootinbib]{revtex4}
\usepackage[dvips]{graphicx}
\usepackage{epsf}
\usepackage{amsmath}
\usepackage{amssymb}

\usepackage{graphicx}
\usepackage{bm}
 \def\gsim{ \lower .75ex \hbox{$\sim$} \llap{\raise .27ex \hbox{$>$}} }
 \def\lsim{ \lower .75ex \hbox{$\sim$} \llap{\raise .27ex \hbox{$<$}} }

\begin{document}

\title{\Large \bf A parametrization for the growth index of linear matter perturbations}
\author{ Puxun Wu$^{1, 2, 3}$,  Hongwei Yu$^{1, 2}$ and Xiangyun Fu$^{1}$,}

\address{$^1$Department of Physics and Institute of  Physics, Hunan
Normal University, Changsha, Hunan 410081, China
\\
$^2$Kavli Institute for Theoretical Physics China, CAS, Beijing
100190, China
\\
$^3$Department of Physics and Tsinghua Center for Astrophysics,
Tsinghua University, Beijing 100084, China }

\begin{abstract}

We propose a parametrization  for the growth index of the linear
matter perturbations, $\gamma(z)=\gamma_0+\frac{z}{1+z}\gamma_1$.
The growth factor  of the  perturbations parameterized as
$\Omega_m^{\gamma}$ is analyzed for both the $w$CDM model and the
DGP model with our proposed form for $\gamma$. We find that
$\gamma_1$ is negative for the $w$CDM model but is positive for the
DGP model. Thus it provides another signature to discriminate them.
We demonstrate that $\Omega_m^{\gamma}$ with $\gamma$ taking our
proposed form approximates the growth factor very well both at low
and high redshfits for both kinds of models. In fact, the error is
below $0.03\%$ for the $\Lambda$CDM model and $0.18\%$ for the DGP
model for all redshifts when $\Omega_{m0}=0.27$. Therefore, our
parametrization may be robustly used to constrain the growth index
of different models with the observational data which include points
for redshifts ranging from 0.15 to 3.8, thus providing
discriminative signatures for different models.
\end{abstract}

\pacs{95.36.+x; 98.80.Es; 04.50.-h}

\maketitle

\section{Introduction}\label{sec1}
The growing observational evidences~\cite{Sne, CMB, SDSS} show that
the present  expansion of our universe is  accelerating. Basically,
two kinds of physical models have been proposed to explain this
mysterious phenomenon. One is dark energy, which has a sufficient
negative pressure to induce a late-time accelerated expansion; the
other is the modified gravity, which originates from the idea that
our understanding to gravity is incorrect in the cosmic scale and
general relativity needs to be modified. However, many different
models proposed so far share the same late time cosmological
expansion, therefore an important task is to discriminate them to
determine which one correctly describes the whole evolution of the
universe. Recently, some attempts have been
made~\cite{r12,r13,r14,r15,r16,r17} in this regard. A particular
effort to discriminate different models focuses on the growth
function $\delta(z)\equiv\delta\rho_m/\rho_m$ of the linear matter
density contrast as a function of redshift
$z$~~\cite{Starobinsky1998, Huterer2007, Sereno2006, Knox2006,
Ishak2006, Acquaviva2008, Daniel2008, Sapone2007, Ballesteros2008,
Bertschinger2008, Laszlo2008, kunz2007, Kiakotou2008,  Linder2007,
Wei2008, Wei20082, Linder2005, Amendola, Nesseris2008, Wang2008,
Boisseau2000, Gong2008, Gong2009, Polarski2008, Gannouji2008,
Gannouji20082, Fu2009}, where $\rho_m$ is the energy density of
matter. While different models give the same late time expansion,
they may produce different growth of matter
perturbations~\cite{Starobinsky1998}.

In order to discriminate different models using the matter
perturbations,  the growth factor $f\equiv \frac{d\ln\delta}{d\ln
a}$ is used. In Ref.~\cite{Fry1985}, the authors found that  this
growth factor can be parameterized  as
 \begin{equation}
 \label{fommegam}
 f=\Omega_m^\gamma,
 \end{equation}
 where $\gamma$ is called the growth
index and   $\Omega_m$ is the fractional energy density of matter.
Using the fact that $\Omega_{m}\simeq 1$ at the high redshift,
treating $\gamma$ as a constant and expanding, around
$\Omega_{m}=1$, the equation obtained by submitting the above
expression into the equation of $f$ (Eq.~(\ref{grwthfeq1}) below),
one can easily obtain the theoretical values of $\gamma$ for
different models. For example, the theoretical values of $\gamma$
for the $\Lambda$CDM model and  the Dvali-Gabadadze-Porrati (DGP)
brane-world model~\cite{Dvali2000} are
$\gamma_\infty=6/11$~\cite{Linder2007, Linder2005} and
$\gamma_\infty =11/16$~\cite{Linder2007, Wei2008}, respectively.
Thus if the value of $\gamma$ can be determined by observations, one
can discriminate these models. Theoretically, if $\gamma$ can be
treated as a constant and the value of $\Omega_{m,0}$ is known, we
can also determine the value of $\gamma$ at $z=0$.  This actually
gives a better approximation to $f$ as will be shown later. However,
the fact that  we do not have a precise value of $\Omega_{m,0}$
 from observations restricts  our ability to obtain an exact value of
constant $\gamma$. Recently by comparing $\Omega^{\gamma_\infty}$
with $f$, the author in Ref.~\cite{Gong2008} found the error is
nearly zero at the high redshift, but  at low redshift,  with
$\Omega_{m,0}=0.27$, the error is larger than $1.2\%$ for
$\Lambda$CDM model and $3\%$ for the DGP model. The discrepancy
originates from the fact that in general $\gamma$ is not a constant
but should be a function of redshift, especially at low redshift
region ($z<2$). Therefore, if one wants to discriminate different
models by using the current observational data on the matter
perturbations with $\gamma$ being taken as a constant, then the
results may be biased, since there are about half of growth factor
data points at the redshift region $z<2$~\cite{Tegmark2006}.

So, it seems necessary to consider an evolutionary growth index
$\gamma(z)$. In this regard  the authors in
Refs.~\cite{Polarski2008, Gannouji2008, Gannouji20082, Fu2009}
studied $\gamma(z)$ with a linear expansion, $\gamma\approx \gamma_0
+ \gamma_0' z$, and found for different models the $\gamma_0'$ is
different, which may provide  another signature to discriminate
different models. Certainly this linear expansion gives a very good
approximation at $z<0.5$, but it is invalid  at high redshift region
and thus is not usable for discriminating different models by
constraining $\gamma_0$ and $\gamma_0'$ from current observations,
since there are few growth factor data points at $z<0.5$. In
Ref.~\cite{Gong2008} the author considered the correction to
$\gamma$ by introducing an $\Omega_m-1$ term and found that, with
$\Omega_{m,0}=0.27$,  the error is blow $0.25\%$ for the
$\Lambda$CDM model and below $0.4\%$  for the DGP model, which are
less than those obtained in the case of a constant
$\gamma=\gamma_\infty$. However, principally speaking,  this
correction cannot be extended to low redshift where the deviation of
$\Omega_{m}$ from $1$ is very large. Therefore, it is desirable to
have a new form of $\gamma(z)$, which is applicable to all the
observational data and can, at the same time, give a very good
approximation to $f$. In this paper we propose a parameterized form
on $\gamma(z)$
 \begin{eqnarray}\label{gammaz}
 \gamma(z)=\gamma_0+\gamma_1 {z\over 1+z}.
 \end{eqnarray}
By numerical calculations, we will demonstrate that  this
parametrization can approximate $f$ very well  at both low and high
redshifts for both the $w$CDM model and the DGP model, and as a
result, it is applicable to all the data points and can be used to
better discriminate these models using observational data.

\section{the $w$CDM model}\label{sec12}
In this section, the dark energy model with a constant equation of
state ($w$CDM) is studied.  To the linear order of matter
perturbations, the growth function $\delta(z)$ at scales much
smaller than the Hubble radius obeys the following equation
\begin{eqnarray}
\label{denpert} \ddot{\delta}+2H\dot{\delta}-4\pi
G_{eff}\,\rho_m\delta=0,
\end{eqnarray}
where  $G_{eff}$ is an effective Newton gravity constant and the dot
denotes the derivative with respect to the time $t$. Using  the
growth factor $f\equiv d\ln\delta/d\ln a$, the above equation
becomes
\begin{eqnarray}
\label{grwthfeq1}
{d\; f\over d\ln
a}+f^2+\left(\frac{\dot{H}}{H^2}+2\right)f=\frac{3}{2}\frac{G_{eff}}{G_{N}}\Omega_m.
\end{eqnarray}

\begin{figure}[htbp]
 \includegraphics[width=0.45\textwidth]{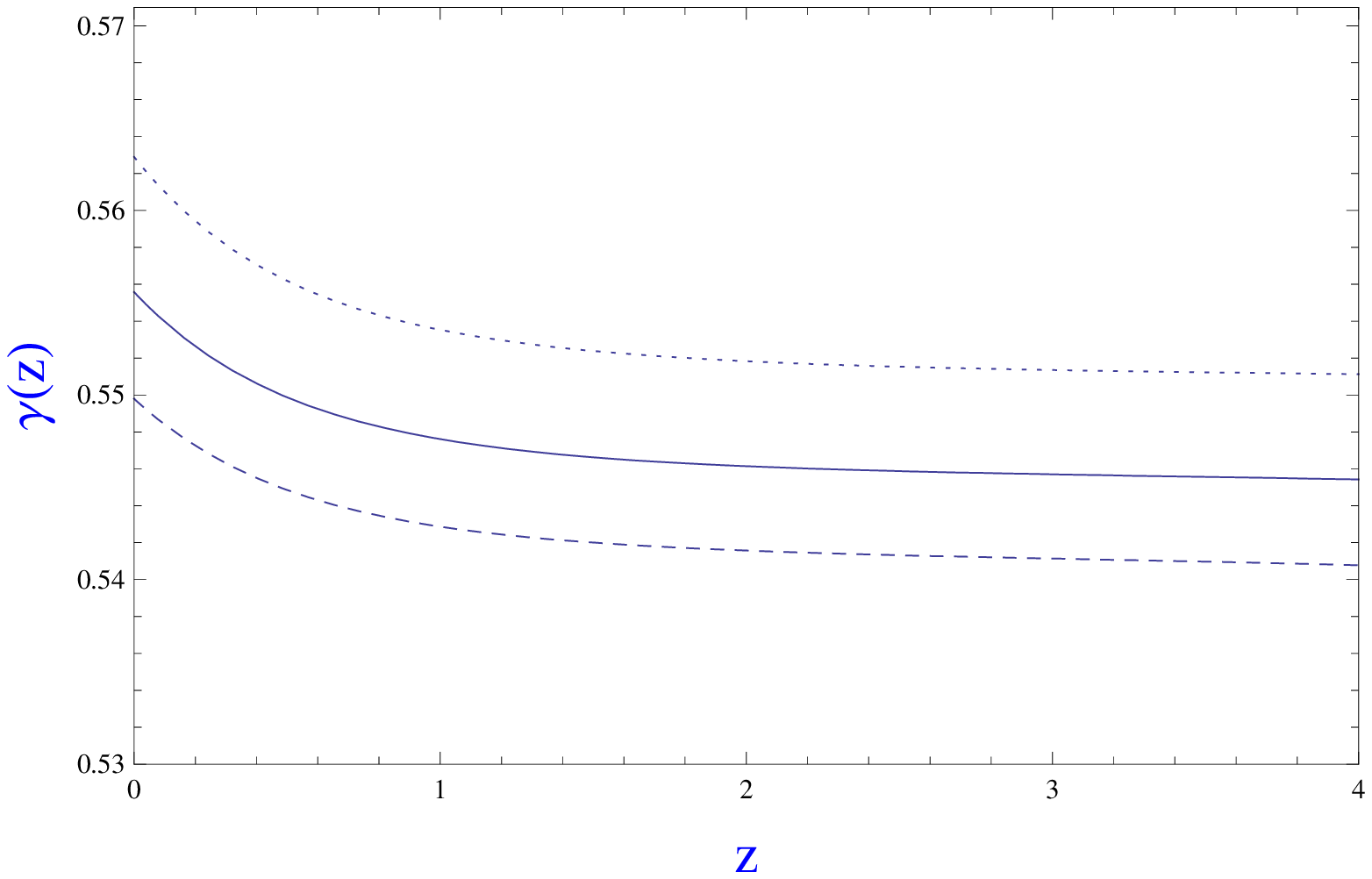}
\caption{\label{fig1} The evolution of growth index $\gamma(z)$ with
redshift  for the $w$CDM model with $\Omega_{m,0}=0.27$. The solid,
dashed and dotted curves correspond to $w=-1$, $-0.8$ and $-1.2$
respectively. }
 \end{figure}

For the $w$CDM dark energy model, the above equation can be
reexpressed as
\begin{eqnarray}
\label{wcdmfeq}
3w\Omega_m(1-\Omega_m)\frac{df}{d\Omega_m}+f^2+\left[\frac{1}{2}
-\frac{3}{2}w(1-\Omega_m)\right]f=\frac{3}{2}\Omega_m\,,
\end{eqnarray}
where ${G_{eff}\over G_{N}}=1$ is used.  In general, it is hard to
obtain an analytical solution to the above equation  and we need to
resort to numerical methods.  In fact, the equation can be solved
numerically by taking into account the condition that $f=1$ at the
high redshift since $\Omega_m=1$ at $z>>1$. Using the
Eq.~(\ref{fommegam}), one can get
\begin{eqnarray}
 \label{gamma-2}
 -(1+z)\gamma^\prime\ln{\Omega_m}+\Omega_m^\gamma+{1\over
 2}[1+3w
 (2\gamma-1)(1-\Omega_m)]
  ={3\over2}\Omega_m^{1-\gamma}\;.
 \end{eqnarray}
This equation is also very hard to be solved analytically.
Fortunately using the relation $f=\Omega_m^{\gamma(z)}$, we can
obtain the evolution of $\gamma(z)$ with the redshift, which is
shown in Fig.~(\ref{fig1}) for $w$CDM model with different values of
$w$. It is easy to see that $\gamma (z)$ can not be regarded as a
constant especially at the redshift region ($z<2$) where some the
observational data points are obtained. Therefore it is obviously
unreliable to discriminate different models with these observational
data while treating the growth index $\gamma$ as a constant.

\begin{figure}[htbp]
 \includegraphics[width=0.45\textwidth]{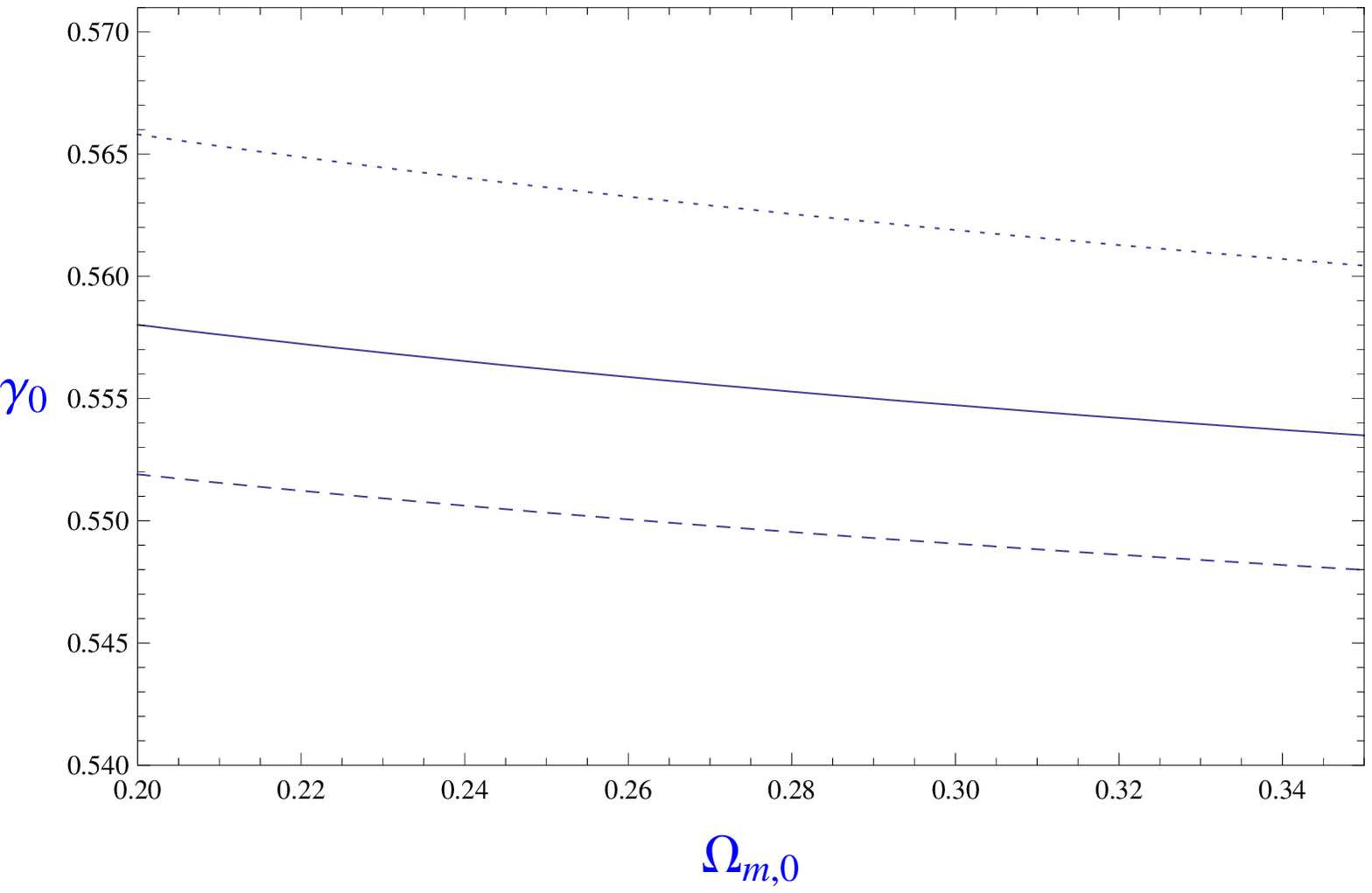}
 \includegraphics[width=0.45\textwidth]{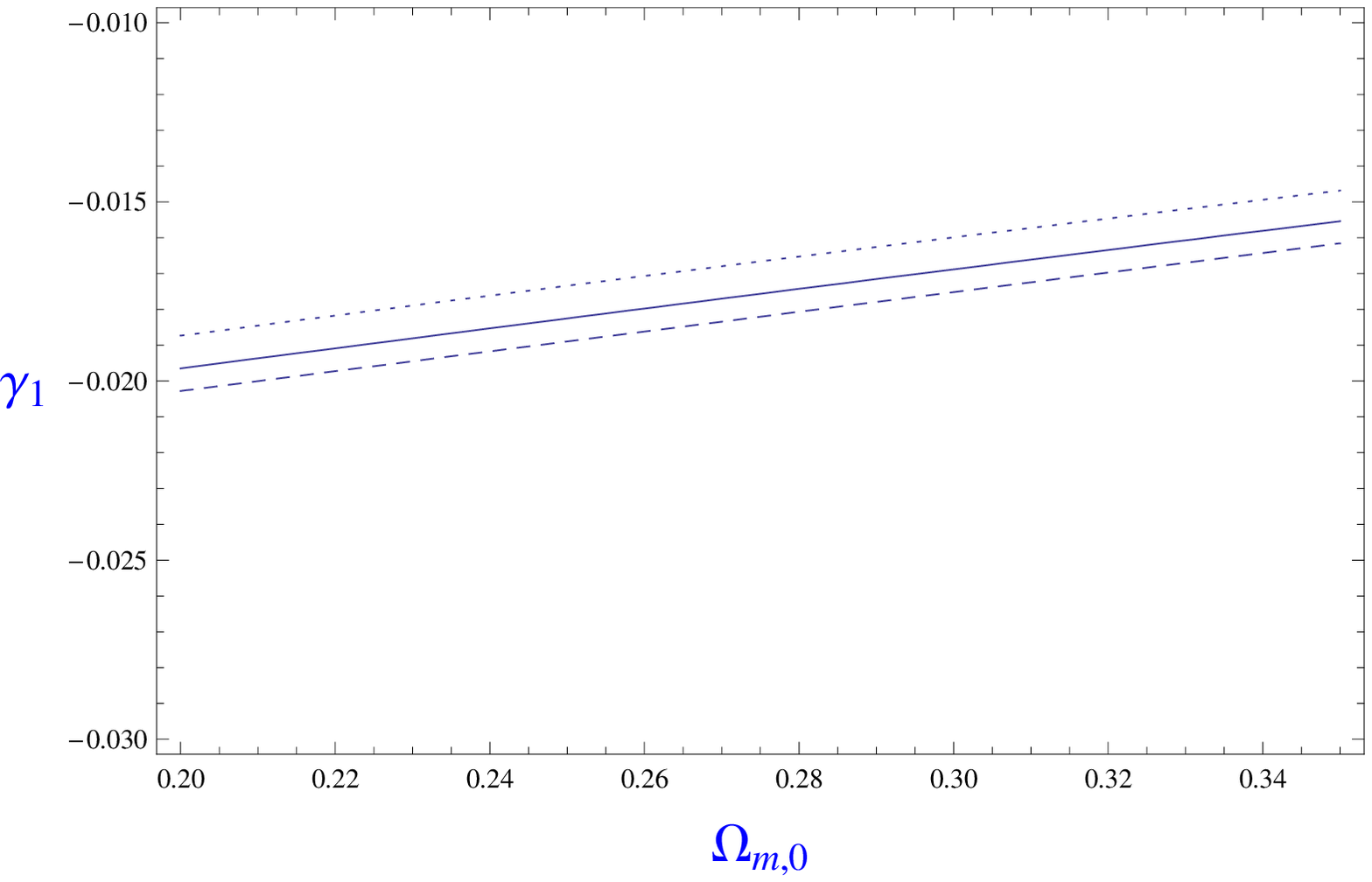}
\caption{\label{fig2} The allowed regions of $\gamma_0$ and
$\gamma_1$ for the   $w$CDM model with $0.2\leq\Omega_{m,0}\leq
0.35$. The solid, dashed and dotted curves correspond to  $w=-1$,
$-0.8$ and $-1.2$ respectively. }
 \end{figure}

Now let us examine if our proposed form of parametrization,
Eq.~(\ref{gammaz}), gives a good approximation to the growth
factor $f$.  If we prior know the value of $\Omega_{m,0}$, and
find the value of $\gamma_0$ through  $f_0$ obtained by
numerically solving solution Eq.~(\ref{wcdmfeq}), then by
substituting Eq.~(\ref{gammaz}) into Eq.~(\ref{gamma-2}), we get
an expression for $\gamma_1$ which yields a different value for a
different redshift $z$. This is because our  proposed
parameterized form  of $\gamma(z)$ is not an exact solution but
only gives an approximation to the curve shown in
Fig.~(\ref{fig1}).  To see how well our approximation is,  let us
take, for simplicity, the value of $\gamma_1$  at the $z=0$,
\begin{eqnarray}
 \label{gamma-3}
 \gamma_1=({\,\ln\Omega_{m,0}^{-1}}\,)^{-1}\bigg[{3\over 2}\Omega_{m,0}^{1-\gamma_0}-\Omega_{m,0}^{\gamma_0}-{3\over
 2}w
 (2\gamma_0-1)(1-\Omega_{m,0})-{1\over2}\bigg]\;,
 \end{eqnarray}
which is determined by the values $\Omega_{m,0}$ and $\gamma_0$.
Since the value of $\Omega_{m,0}$ can be determined by the
observations and $\gamma_0$ can be obtained by the numerical
solution of $f$ with the relation $f_0=\Omega_{m,0}^{\gamma_0}$,
we can get the value of $\gamma_1$. In Fig.~(\ref{fig2}), we show
the allowed region of $\gamma_0$ and the corresponding $\gamma_1$
with a prior given region of $\Omega_{m,0}$:
$0.20\leq\Omega_{m,0}\leq0.35$. From this Figure we can see that
$\gamma_0$ and $\gamma_1$ vary only slightly as $\Omega_{m,0}$
changes.

\begin{figure}[htbp]
 \includegraphics[width=0.55\textwidth]{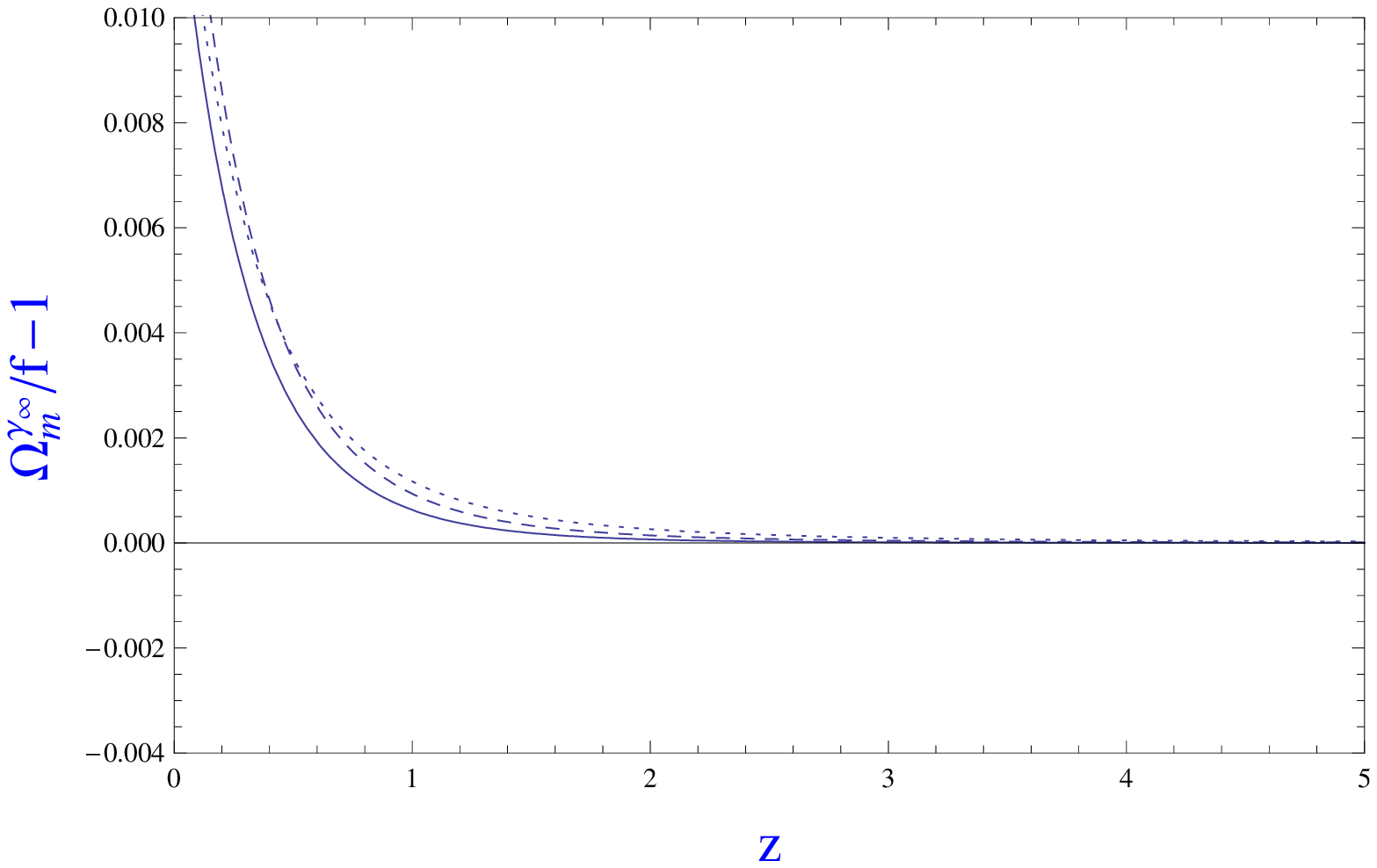}
 \includegraphics[width=0.55\textwidth]{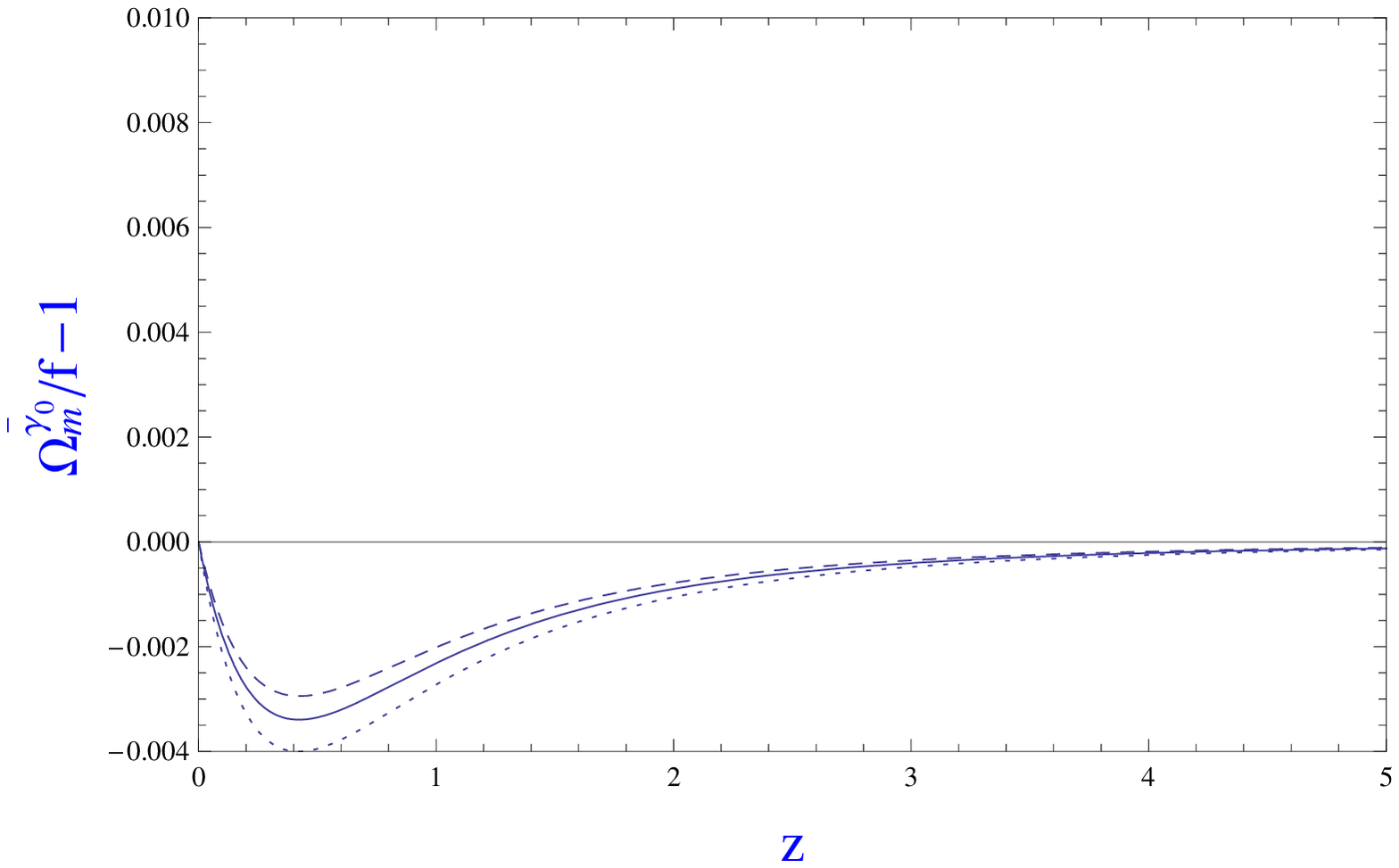}
 \includegraphics[width=0.55\textwidth]{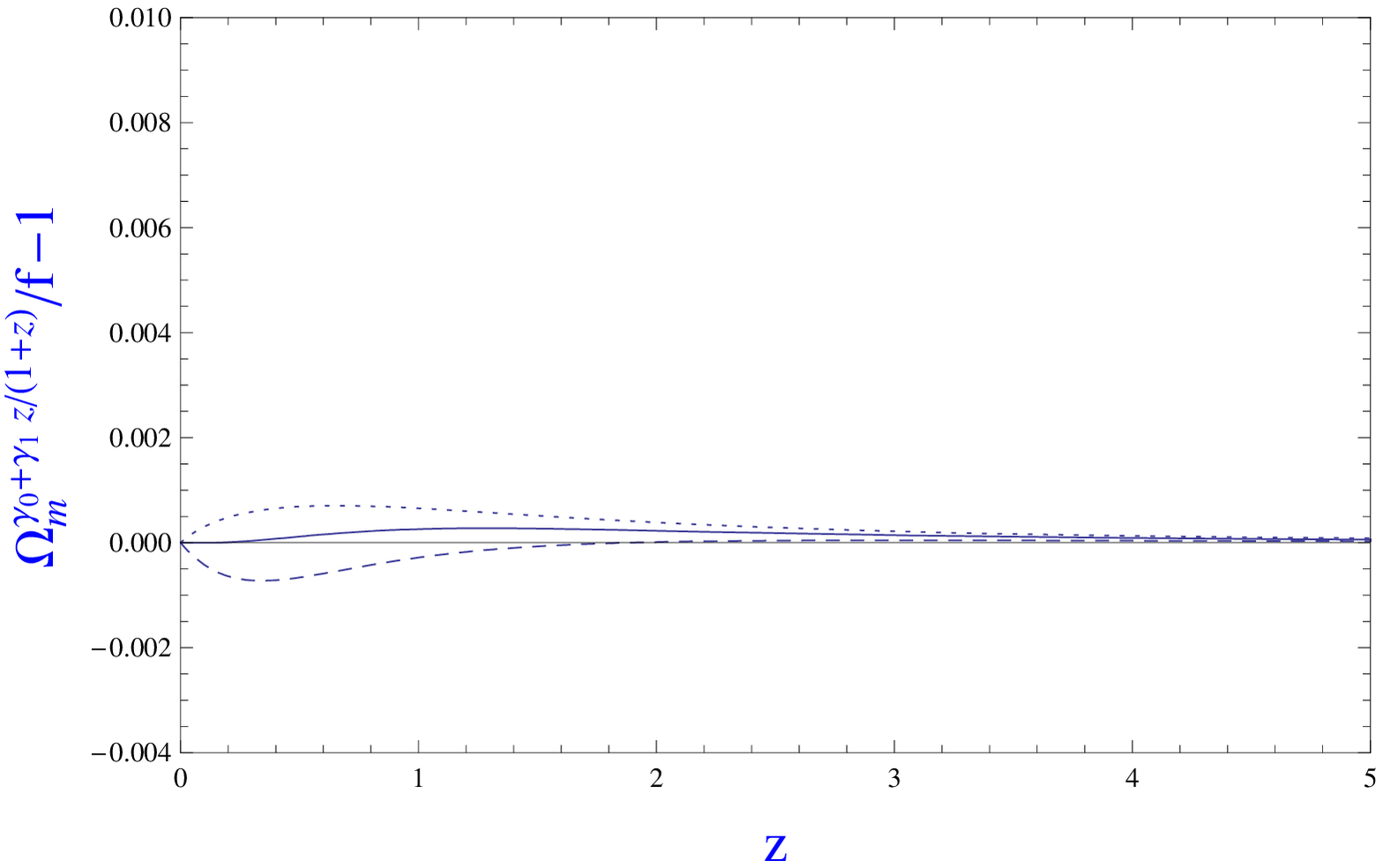}
\caption{\label{fig3}
 The relative difference between the growth factor $f$ and
$\Omega_m^{\gamma}$ for the $w$CDM model with $\Omega_{m,0}=0.27$.
The upper, middle and bottom  panels show the results of
$\gamma=\gamma_\infty$, $\overline{\gamma}_0$  and
$\gamma=\gamma_0+z/(1+z) \gamma_1$, respectively. The solid, dashed,
and dotted curves show the results  for $w=-1$, $-0.8$ and $-1.2$,
respectively.}
 \end{figure}

Now we discuss how well $\Omega_m^{\gamma}$, with $\gamma$ taking
our parameterized form, approximates the growth factor $f$.
Numerical results are shown in Fig.~(\ref{fig3}). Three different
cases are plotted for comparison. One is that the growth index is
treated as a constant which is determined at very high redshift
where $\Omega_m=1$ ($\gamma_\infty$). This case is shown on the
upper panel. The second is that $\gamma$ is also treated as a
constant but the constant is obtained at $z=0$ denoted as
$\overline{\gamma}_0$ with $\overline{\gamma}_0=\ln f(0)/\ln
\Omega_{m,0}(0)$. The result is shown in the middle panel. From the
Figure, one can see that $\Omega_m^{\gamma}$ with a constant
$\gamma$ given by $\overline{\gamma}_0$  approximates $f$ better at
low redshifts than that given by $\gamma_\infty$, but is not as good
at high redshifts ($z>1$). This is because $\overline{\gamma}_0$ is
a good approximation at the low redshift while $\gamma_\infty$ gives
a good approximation at the high redshift. The third is that
$\gamma$ is evolving with redshift and takes our proposed form given
in Eq.~(\ref{gammaz}). The result is shown in the bottom  panel. Now
$\Omega_m^{\gamma}$ approximates $f$ very well both at low and high
redshfits, and remarkably it approximates $f$ better than $0.08\%$
even at low redshfits. Notice, however, that for
$\Omega_m^{\gamma_{\infty}}$ the error is over $1\%$ at low
redshifts and  for $\Omega_m^{\overline{\gamma}_0}$ the largest
error is $0.4\%$. For the case of $w=-1$ ($\Lambda$CDM), we find
that the error resulting from using our proposed form is below
$0.03\%$. This is much less than that, obtained in
Ref.~\cite{Gong2008}, for $\Omega_m^{\gamma}$  with corrections to a
constant $\gamma_\infty $ added through expanding  $\gamma$  at
$\Omega_m=1$ as $\gamma=\gamma_\infty+\frac{15}{1331}(1-\Omega_m)$,
where  the error is only blow $0.25\%$.  We get approximately one
order of magnitude improvement in terms of errors in the
approximation. Therefore, the form given in Eq.~(\ref{gammaz}) is
basically very close to the real evolution of $\gamma$ with
redshift.

\section{the DGP model}\label{sec2}
\begin{figure}[htbp]
 \includegraphics[width=0.45\textwidth]{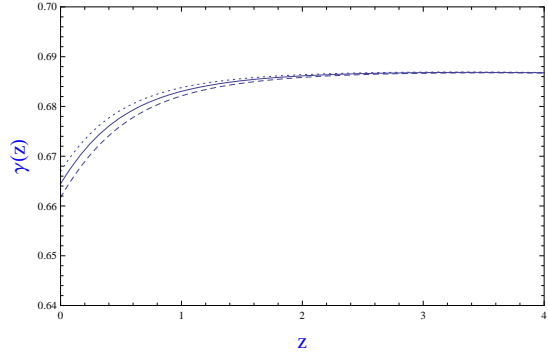}
\caption{\label{fig4} The evolution of growth index $\gamma(z)$ with
the redshift  for the DGP model.  The solid, dashed and dotted
curves correspond to $\Omega_{m,0}=0.27$, $0.24$ and  $0.30$
respectively. }
 \end{figure}

\begin{figure}[htbp]
 \includegraphics[width=0.45\textwidth]{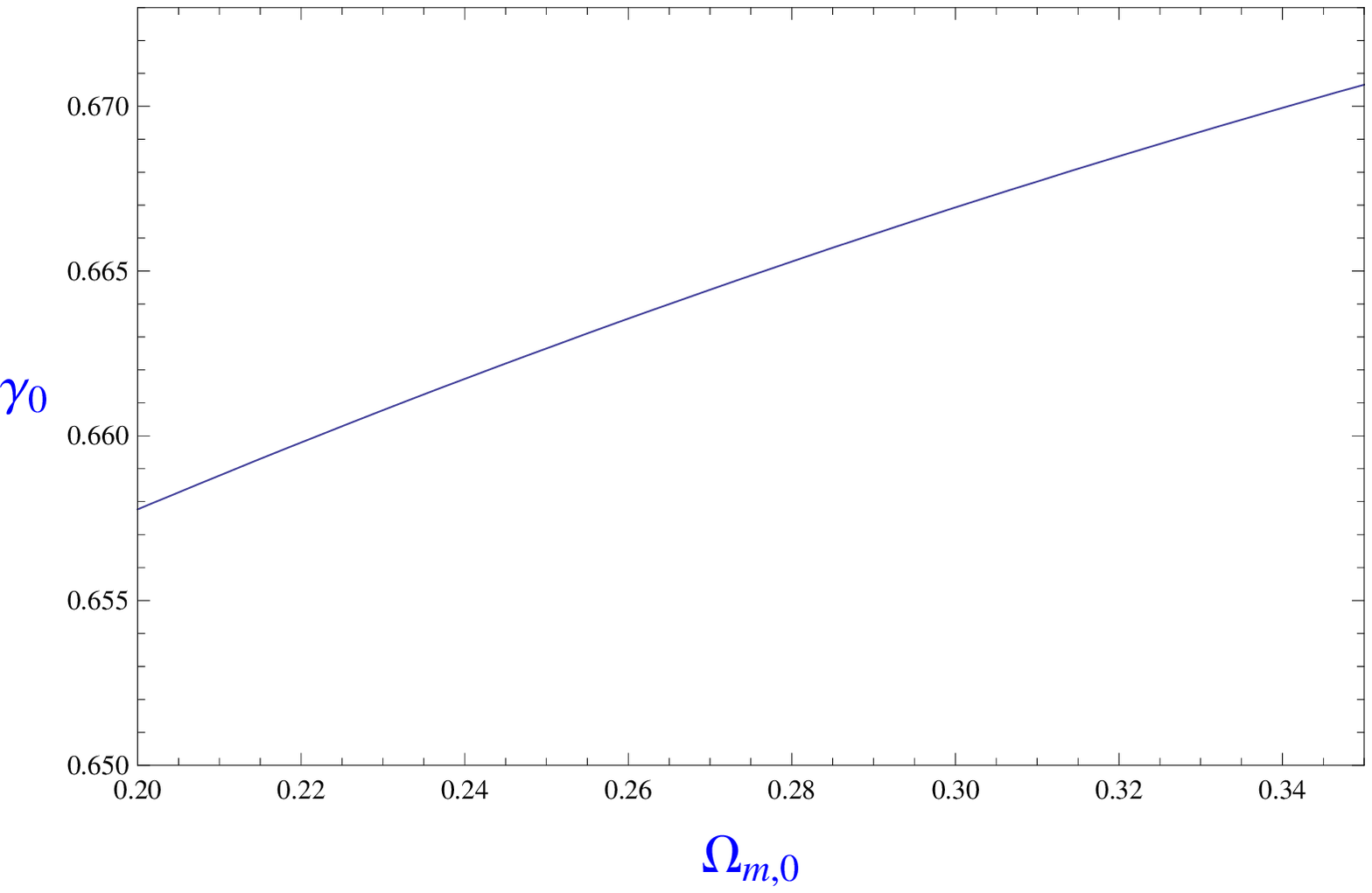}
 \includegraphics[width=0.45\textwidth]{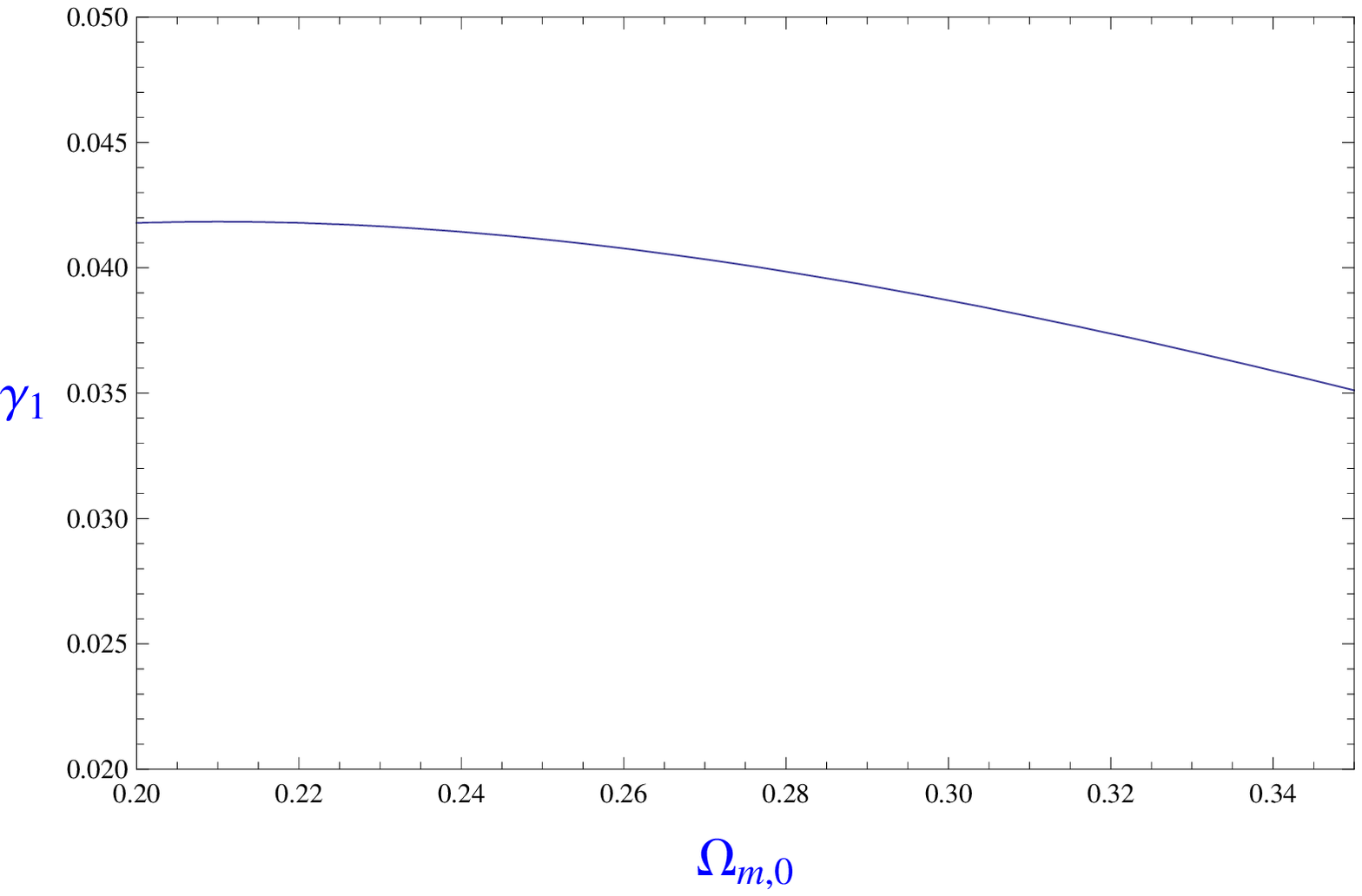}
\caption{\label{fig5} The allowed regions of $\gamma_0$ and
$\gamma_1$ for the DGP model with   $0.2\leq\Omega_{m,0}\leq 0.35$.
}
 \end{figure}

Now we will discuss a modified gravity model, the DGP model. For
this model, the effective Newton constant takes the following form
 \begin{eqnarray}\label{eq12}
 {G_{\rm eff}\over G_{N}}= \frac{2(1+2\Omega_m)}{3(1+\Omega_m^2)}.
 \end{eqnarray}
According to Refs.~\cite{Wei2008, Gong2009, Lue2004, Koyma2006}, the
growth factor $f$ satisfies the equation
\begin{eqnarray}
-3\frac{(1-\Omega_m)\Omega_m}{1+\Omega_m} {df\over
 d\Omega_m}+f^2+f\frac{2-\Omega_m}{1+\Omega_m}
 =\frac{\Omega_m(1+2\Omega_m^2)}{1+\Omega_m^2}.
\label{omegadgpdf}
 \end{eqnarray}
Submitting Eq.~(\ref{fommegam}) into the above equation, we can
obtain an equation of $\gamma (z)$
 \begin{eqnarray}
  &&{1\over
2}\bigg[1- \frac{3(1-\Omega_m)}{1+\Omega_m} (2\gamma-1)\bigg]
-(1+z)\gamma^\prime\ln{\Omega_m}+\Omega_m^\gamma=\frac{1+2\Omega_m^2}{1+\Omega_m^2}\Omega_m^{1-\gamma}\,.
 \end{eqnarray}
The numerical solution for $\gamma(z)$ is shown in Fig.~(\ref{fig4})
with different values of $\Omega_{m,0}$. From Figure, one can see
that the $\gamma(z)$ is evolutionary especially at low redshifts
($z<2$) and is an increasing function of redshift, in contrast to
the $w$CDM model, where $\gamma(z)$ is a decreasing one. Now, one
can show that $\gamma_1$  determined at $z=0$ is given
 \begin{eqnarray}
 \label{gamma0prime-1}
  \gamma_1&=&(\ln\Omega_{m,0}^{-1})^{-1}\bigg[-\Omega_{m,0}^{\gamma_0}
  +\frac{1+2\Omega_{m,0}^2}{1+\Omega_{m,0}^2}\Omega_{m,0}^{1-\gamma_0}-{1\over
  2}+\frac{3(1-\Omega_{m,0})}{1+\Omega_{m,0}}
  (\gamma_0-{1 \over
  2})\bigg]\,.
 \end{eqnarray}
In Fig.~(\ref{fig5}) we plot the possible region of $\gamma_0$ and
$\gamma_1$  for $0.20\leq\Omega_{m,0}\leq0.35$. It is easy to see
that $\gamma_0$ increases from $0.658$ to $0.671$ and $\gamma_1$
decreases from $0.042$ to $0.035$. Apparently  $\gamma_1$ here is
positive, whereas $\gamma_1(\approx-0.02)$ in the $w$CDM model is
negative. Therefore, $\gamma_1$  also provides a signature to
discriminate the DGP model and the $w$CDM model.

\begin{figure}[htbp]
 \includegraphics[width=0.55\textwidth]{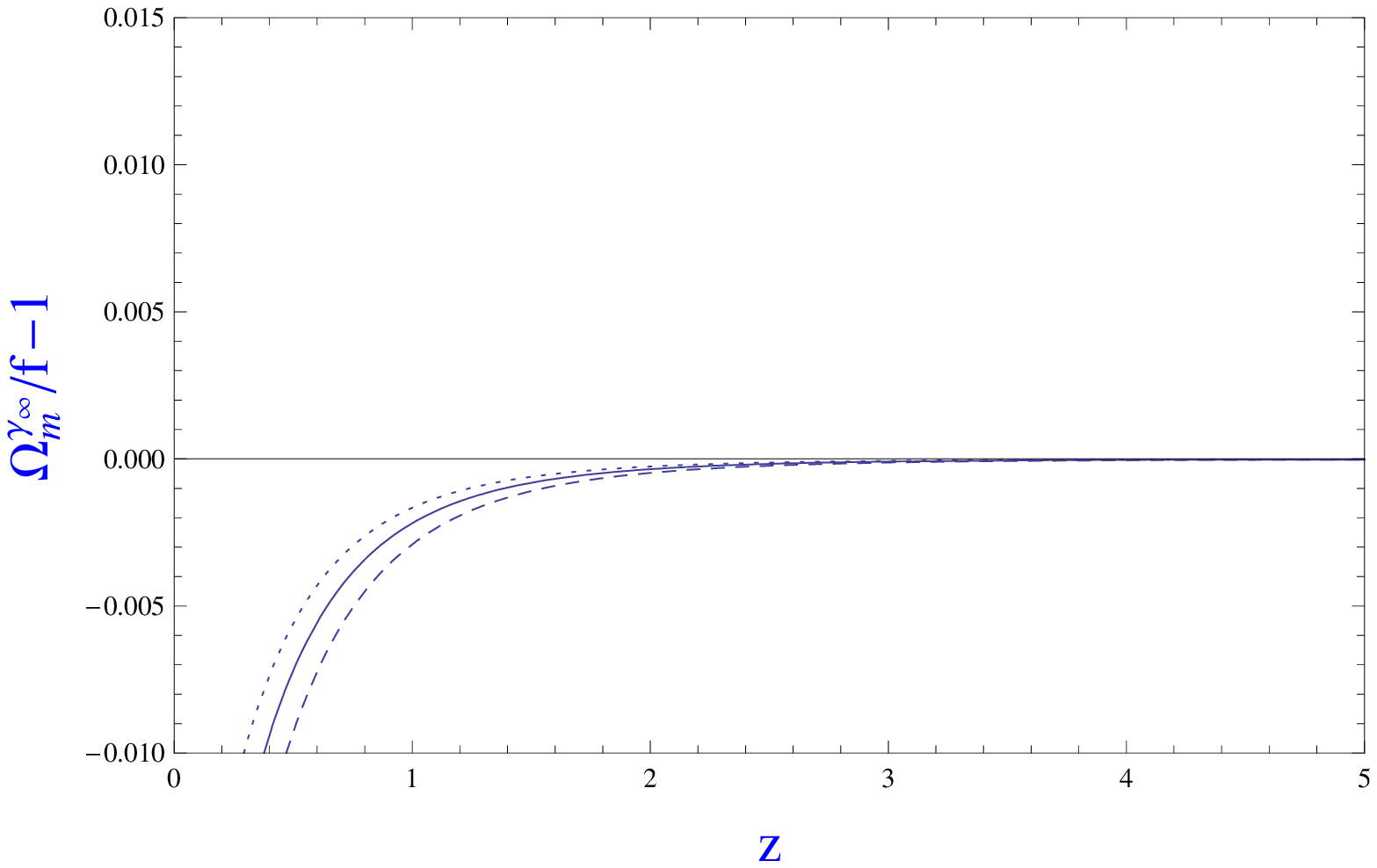}
 \includegraphics[width=0.55\textwidth]{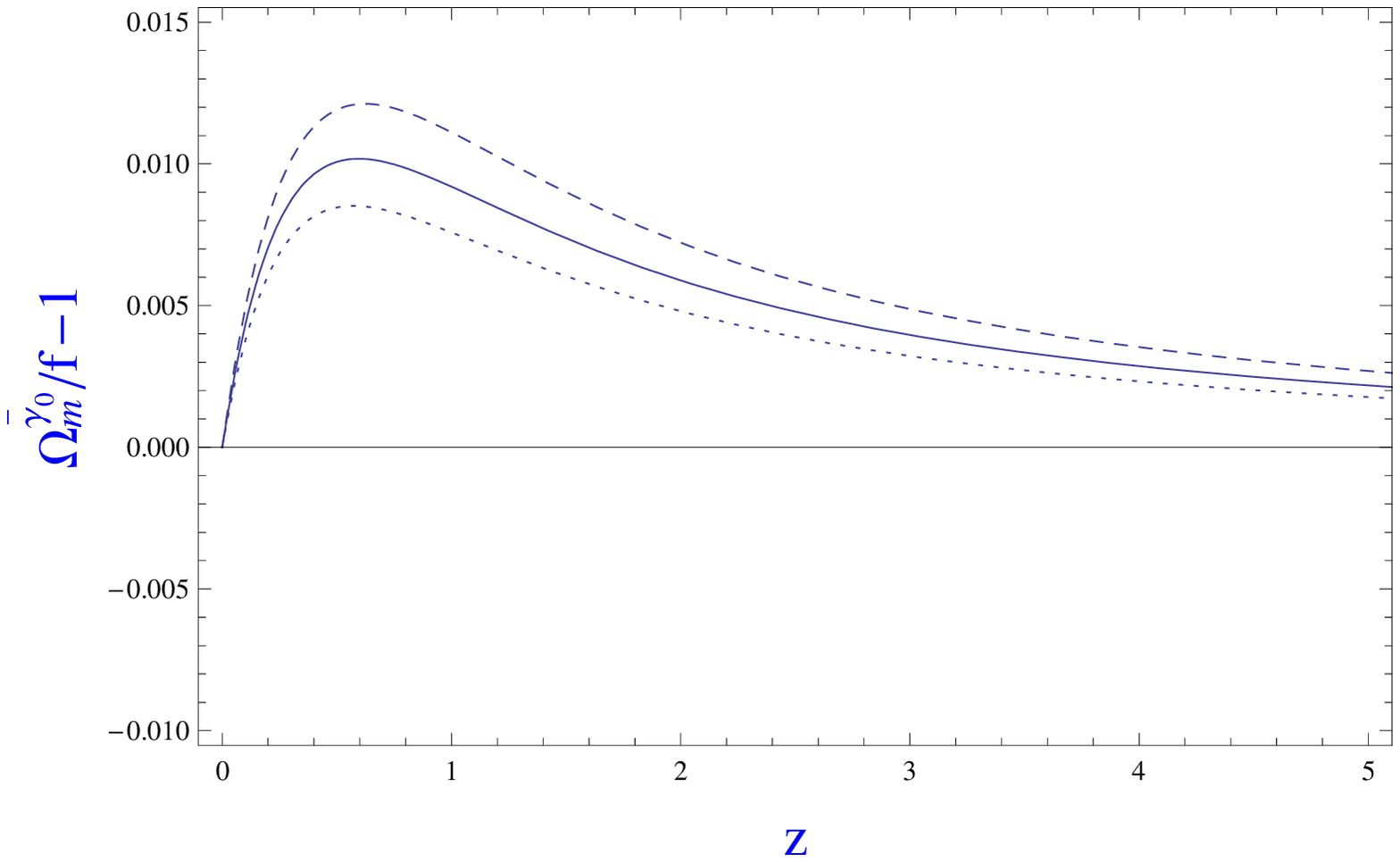}
 \includegraphics[width=0.55\textwidth]{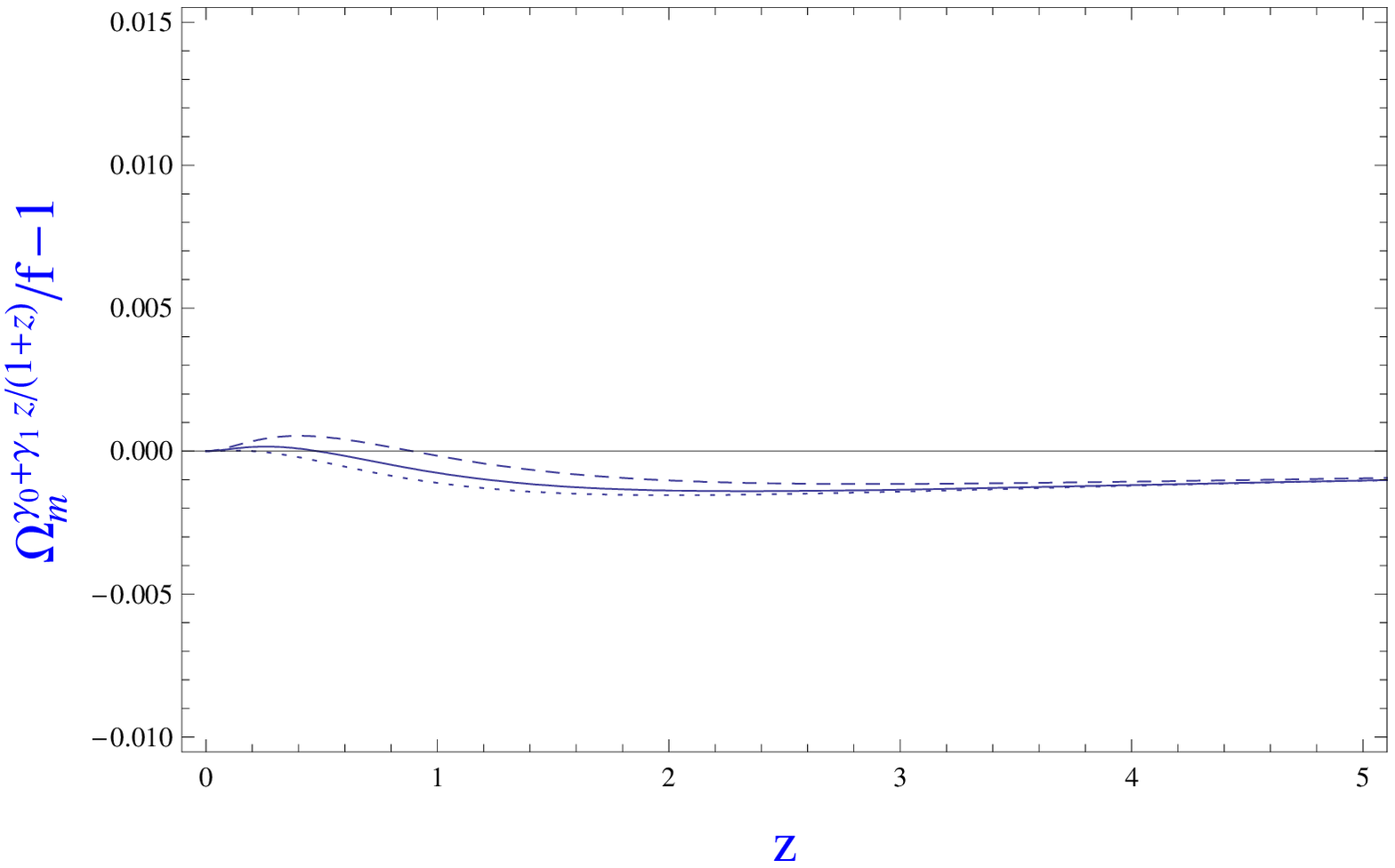}
\caption{\label{fig6}
 The relative difference between the growth factor $f$ and
$\Omega_m^{\gamma}$ for the DGP model.  The upper, middle and bottom
panels show the results of $\gamma=\gamma_\infty$,
$\overline{\gamma}_0$  and $\gamma=\gamma_0+z/(1+z) \gamma_1$,
respectively. The solid, dashed, and dotted curves show the results
for $\Omega_{m,0}=0.27$,  $0.24$ and $0.30$, respectively.}
 \end{figure}

How well the $\Omega_m^{\gamma}$ approximate the growth factor $f$
is  discussed  as done in above section and the results are shown in
Fig.~(\ref{fig6}). Comparing the upper and middle panels of this
figure, we find, as in the $w$CDM model,  that $\Omega_m^{\gamma}$
with constant  $\gamma$ given by $\overline{\gamma}_0$ approximates
$f$ better at low redshifts than that given by $\gamma_\infty$, but
is not as good at high redshifts ($z>1$). Comparing these two panels
with  the bottom  one, one sees that at low redshifts  $\gamma(z)$
proposed in the present paper gives the best approximation. At low
redshifts, the error using our proposed form for $\gamma(z)$  is
blow $0.2\%$ , but for a constant $\gamma$, the error is larger than
$2\%$ when $\gamma=\gamma_{\infty}$ and is only blow $1.2\%$ when
$\gamma=\overline{\gamma}_0$. At low redshifts, the approximation
with our proposed form is also better than that obtained in
Ref.~\cite{Gong2008} where the constant $\gamma$ is corrected as
$\gamma=\gamma_\infty+\frac{7}{5632}(1-\Omega_m)$, since, when
$\Omega_{m,0}=0.27$, the error of our proposed form is blow $0.18\%$
while it is only blow $0.25\%$ for  $\gamma$ corrected with a
$1-\Omega_m$ term.  It is  worthy to note that
$\Omega_m^{\gamma(z)}$ with $\gamma(z)$ taking our proposed form
still approximates $f$ very well for all redshifts,  since the
largest error is only $0.18\%$, when $\Omega_{m,0}=0.27$.

\section{Conclusion}\label{sec4}

In this paper, we propose a parameterized form for the growth index
of the linear matter perturbations,
$\gamma(z)=\gamma_0+\frac{z}{1+z}\gamma_1$. The growth factor  of
the linear matter perturbations  is analyzed for both the $w$CDM
model and the DGP model. We find that $\gamma_1$ is negative for the
$w$CDM model  but  is positive for the DGP model. Thus it provides
another signature to discriminate them. If we parameterize the
growth factor $f$ as $\Omega_m^{\gamma}$, then at low redshifts,
$\Omega_m^{\gamma}$ with $\gamma$ taking our proposed form
approximates the growth factor $f$ better than that in the case of a
constant $\gamma$ with or without the $1-\Omega_m$ correction term.
At high redshifts,  the approximation is also very good.  In fact,
the error is below $0.03\%$ for the $\Lambda$CDM model and $0.18\%$
for the DGP model for all redshifts when $\Omega_{m0}=0.27$.
Therefore, our parametrization can be robustly  used to constrain
the growth index of different models with the observational data
which include points for redshifts ranging from 0.15 to 3.8,  thus
providing discriminative signatures for different models.

\section*{Acknowledgments}
This work was supported in part by the National Natural Science
Foundation of China under Grants No.10775050, 10705055, the SRFDP
under Grant No. 20070542002, the Research Fund of Hunan Provincial
Education Department,  the Hunan Provincial Natural Science
Foundation of China under Grant No. 08JJ4001, and the China
Postdoctoral Science Foundation.

\begin{thebibliography}{99}

\bibitem{Sne}A. G. Riess, A. V. Filippenko, P. Challis, {\it et al.}, Astron. J. {\bf 116}, 1009 (1998);
          S. J. Perlmutter, G. Aldering, G. Goldhaber, {\it et al.}, Astrophy. J. {\bf 517}, 565 (1999);
         J.~L.~Tonry {\it et al.}, Astrophys. J.  {\bf 594}, 1 (2003);
         R.~A.~Knop {\it et al.}, Astrophys. J.  {\bf 598}, 102 (2003);
         A.~G.~Riess {\it et al.}, Astrophys. J.  {\bf 607}, 665 (2004);
         A.~G.~Riess {\it et al.}, Astrophys. J.  {\bf 659}, 98 (2007);
         P.~Astier {\it et al.}, Astron. Astrophys.  {\bf 447}, 31 (2006);
         J.~D.~Neill {\it et al.}, Astron.\ J.\  {\bf 132}, 1126 (2006);
         W. M. Wood-Vasey {\it et al.}, Astrophys. J. {\bf 666},  694
         (2007);
         T. M. Davis {\it et al.}, Astrophys. J. {\bf 666},  716 (2007).
\bibitem{CMB}
         D. N. Spergel, et al., Astrophys. J. Suppl. {\bf 170}, 377 (2007);
         L. Page, et al., Astrophys. J. Suppl. {\bf 170}, 335 (2007);
         G. Hinshaw, et al., Astrophys. J. Suppl. {\bf 170}, 288 (2007);
         N. Jarosik, et al., Astrophys. J. Suppl. {\bf 170}, 263 (2007).

\bibitem{SDSS}
               D. J. Eisenstein, {\it et al.}, Astorphys. J. {\bf 633},  560  (2005);
               E. Komatsu, {\it et al.}, Astrophys. J. Suppl. {\bf 180}, 330 (2009).

\bibitem{r12}
            R. R. Caldwell and E. V. Linder, Phys. Rev. Lett.  {\bf 95}, 141301 (2005);
            E. V. Linder, Phys. Rev.  D {\bf 73}, 063010 (2006).

\bibitem{r13}
           R. J. Scherrer,  Phys. Rev.  D {\bf 73}, 043502 (2006).

\bibitem{r14}
           T. Chiba, Phys. Rev.  D {\bf 73}, 063501 (2006).

\bibitem{r15}
           E. V. Linder, Gen. Rel. Grav.  {\bf 40}, 329 (2008).

\bibitem{r16}
            V. Sahni, T. D. Saini, A. A. Starobinsky and U. Alam, JETP Lett. {\bf 77}, 201
            (2003);
            U. Alam, V. Sahni, T. D. Saini and A. A. Starobinsky, Mon.  Not. Roy. Astron. Soc.  {\bf 344}, 1057 (2003).


\bibitem{r17}
            H. Wei and R. G. Cai,  Phys. Lett. B {\bf 655}, 1 (2007).



\bibitem{Starobinsky1998}A. A. Starobinsky, JETP Lett. {\bf 68}, 757 (1998).


\bibitem{Huterer2007}D. Huterer and E. V. Linder, Phys. Rev. D {\bf 75}, 023519 (2007).
\bibitem{Sereno2006} M. Sereno and J. A. Peacock, Mon. Not. Roy. Astron. Soc. {\bf 371}, 719 (2006).
\bibitem{Knox2006}   L. Knox, Y.-S. Song and J. A. Tyson, Phys. Rev. D {\bf 74}, 023512 (2006).
\bibitem{Ishak2006}  M. Ishak, A. Upadhye and D. N. Spergel, Phys. Rev. D 74, 043513 (2006).
\bibitem{Acquaviva2008} V. Acquaviva, A. Hajian, D. N. Spergel and S. Das, Phys. Rev. D {\bf 78}, 043514 (2008).
\bibitem{Daniel2008}T. Koivisto and D. F. Mota, Phys. Rev. D {\bf 73} (2006) 083502;
D. F. Mota, J. R. Kristiansen, T. Koivisto and N. E. Groeneboom£¬
Mon.Not.Roy.Astron.Soc. {\bf 382} (2007) 793; S. Daniel, R.
Caldwell, A. Cooray and A. Melchiorri, Phys. Rev. D {\bf 77}, 103513
(2008).
\bibitem{Sapone2007} D. Sapone and L. Amendola, arXiv: 0709.2792.
\bibitem{Ballesteros2008}G. Ballesteros and A. Riotto, Phys. Lett. B {\bf 668}, 171 (2008).
\bibitem{Bertschinger2008} E. Bertschinger and P. Zukin, Phys. Rev. D {\bf 78}, 024015 (2008).
\bibitem{Laszlo2008} I. Laszlo and R. Bean, Phys. Rev. D {\bf 77}, 024048 (2008).
\bibitem{kunz2007} M. Kunz and D. Sapone, Phys. Rev. Lett. {\bf 98}, 121301 (2007).
\bibitem{Kiakotou2008} A. Kiakotou, O Elgaro and O. Lahav, Phys. Rev. D {\bf 77}, 063005 (2008).
\bibitem{Linder2007}
             E. V. Linder and R. N. Cahn, Astropart.  Phys. {\bf 28}, 481 (2007).

\bibitem{Wei2008} H. Wei, Phys. Lett. B {\bf 664}, 1  (2008).
\bibitem{Wei20082}H. Wei, S. N. Zhang, Phys. Rev. D {\bf 78}, 023011 (2008).

\bibitem{Gong2008}Y.  Gong, Phys. Rev. D {\bf 78}, 123010 (2008).
\bibitem{Gong2009} Y. Gong, M. Ishak and A. Wang, arXiv: 0903.0001.

\bibitem{Linder2005}E. V. Linder, Phys.  Rev.   D {\bf 72}, 043529 (2005).



\bibitem{Amendola}
         C. Di Porto and L. Amendola, Phys. Rev. D {\bf 77}, 083508 (2008); 
         L. Amendola, M. Kunz and D. Sapone,  arXiv: 0704.2421.
\bibitem{Nesseris2008}
         S. Nesseris and L. Perivolaropoulos, Phys. Rev.  D {\bf 77}, 023504 (2008). 

\bibitem{Wang2008}
          Y. Wang, J. Cosmol. Astropart. Phys. {\bf 5}, 21 (2008).


\bibitem{Boisseau2000}
          B. Boisseau, G. Esposito-Far¨¨se, D. Polarski, A. A. Starobinsky, Phys. Rev. Lett. {\bf 85},  2236 (2000);
          M.J. Mortonson, W. Hu and D. Huterer, Phys. Rev. D 79 (2009) 023004;
 J. He, B. Wang, Y. P. Jing, arXiv:0902.0660; J. B. Dent,
S. Dutta and L. Perivolaropoulos, arXiv:0903.5296; M. Ishak and J.
Dossett, arXiv:0905.2470.

\bibitem{Polarski2008}
         D. Polarski and R. Gannouji, Phys. Lett.  B {\bf 660}, 439
         (2008).
\bibitem{Gannouji2008}
         R. Gannouji and D. Polarski, J. Cosmol. Astropart. Phys. {\bf 05}, 018 (2008).
\bibitem{Gannouji20082}
         R. Gannouji, B. Moraes and D. Polarski, arXiv: 0809.3374.
\bibitem{Fu2009}X. Fu, P. Wu and H. Yu, Physics Letters B 677 (2009) 12

\bibitem{Fry1985}J. N. Fry, Phys. Lett. B {\bf 158}, 211 (1985);
                 A. P. Lightman and P. L. Schechter, Astrophys. J. {\bf 74}, 831 (1990);
                 L. Wang and P. J. Steinhardt, Astrophys. J. {\bf 508}, 483 (1998);


\bibitem{Dvali2000} G. Dvali, G. Gabadadze and M. Porrati, Phys. Lett. B {\bf 485}, 208 (2000);
                    Z. Zhu, M. Sereno, Astro. Astrophys. {\bf 487},   831 (2008).

\bibitem{Tegmark2006} L. Guzzo et al., Nature {\bf 451}, 541 (2008).
                      M. Tegmark el al., Phys. Rev. D {\bf 74}, 123507 (2006);
                      N. P. Ross et al., Mont. Not. R. Astron. Soc. {\bf 381}, 573 (2007);
                      J. da Angela et al., Mont. Not. R. Astron. Soc. {\bf 383}, 565 (2008);
                      P. McDonald et al., Astrophys. J. {\bf 635}, 761 (2005);
                      M. Viel, M. G. Haehnelt and V. Springel, Mont. Not. R. Astron. Soc. {\bf 354}, 684 (2004);
                      M. Viel, M. G. Haehnelt and V. Springel, Mont. Not. R. Astron. Soc. {\bf 365}, 231 (2006).

\bibitem{Lue2004}A. Lue, R. Scoccimarro and G. D. Starkman, Phys. Rev. D {\bf 69}, 124015 (2004).
\bibitem{Koyma2006} K. Koyama and R. Maartens, J. Cosmol. Astropart. Phys. {\bf 01}, 016 (2006).

\end{thebibliography}
\end{document}